\documentclass[twocolumn,preprintnumbers,amsmath,superscriptaddress,amssymb,aps]{revtex4-1}

\usepackage{graphicx}
\usepackage{dcolumn}
\usepackage{bm}
\usepackage{amsmath}
\usepackage{amssymb}
\usepackage{graphicx}
\usepackage{indentfirst}
\usepackage{booktabs}
\usepackage{multirow}
\usepackage{colortbl}

\selectfont

\begin{document}
\title{Two-dimensional transition metal oxides Mn$_2$O$_3$ realized quantum anomalous Hall effect}

\author{Ping Li}
\address{School of Physical Science and Technology, Soochow University, Suzhou 215006, People's Republic of China}
\author{Tian-Yi Cai}
\email{caitianyi@suda.edu.cn}
\address{School of Physical Science and Technology, Soochow University, Suzhou 215006, People's Republic of China}

\date{\today}

\begin{abstract}
The quantum anomalous Hall effect is an intriguing topological nontrivial phase arising from spontaneous magnetization and spin-orbit coupling. However, the tremendously harsh realizing requirements of the quantum anomalous Hall effects in magnetic topological insulators of Cr or V-doped (Bi,Sb)$_2$Te$_3$ film, hinder its practical applications. Here, we perform first-principle calculations to predict that three Mn$_2$O$_3$ structure are intrinsic ferromagnetic Chern insulators. Remarkably, a quantum anomalous Hall phase of Chern number C = -2 is found, and there are two corresponding gapless chiral edge states appearing inside the bulk gap. More interestingly, only a small tensile strain is needed to induce the phase transition from $\emph{Cmm2}$ and $\emph{C222}$ phase to $\emph{P6/mmm}$ phase. Meanwhile, a topological quantum phase transition from a quantum anomalous Hall phase to a trivial insulating phase can be realized. The combination of these novel properties renders the two-dimensional ferromagnet a promising platform for high efficient electronic and spintronic devices.
\end{abstract}

\maketitle

\section{Introduction}
In the last few decades, two-dimensional (2D) materials have attracted enormous attention due to their exotic physical phenomena and unique properties contrasting to their bulk counterparts materials. \cite{1,2,3,4} Numerous investigations have been focused on pristine 2D materials. Their intrinsic magnetic is found to be controlled by the Mermin-Wagner theorem, \cite{5} which describes the dominant effects of additional thermal fluctuations in the limited phase space of a low-dimensional system. This restriction immensely impedes the development of these 2D nanomaterials in the field of nanoelectronics and nanospintronics. For this purpose, 2D magnetic materials with fascinating electronic and magnetic properties are greatly needed.

Recently, Cr$_2$Ge$_2$Te$_6$ and CrI$_3$ are the ultrathin laminae exfoliation from intrinsically ferromagnetic (FM) vdW materials, whose magnetic behaviour has been investigated down to a monolayer thickness. \cite{6,7} Hence, the successful preparation of 2D magnetic materials provides a platform to explore intriguing properties, especially the quantum anomalous Hall (QAH) state. It is insulating in the bulk but exhibits robust chiral edge states, leading to quantized Hall conductivity in absence of an external magnetic field. \cite{8,9} The synergistic effect between the magnetic material which broken time reversal symmetry (TRS) and the spin-orbit coupling (SOC) could lead to an insulating state with a nonzero Chern number. Due to the nontrivial topological properties and intriguing potential application for designing low-energy consumption electronics and spintronic devices, enormous theoretical investigations have been made recently to search for real QAH insulators. \cite{10,11,12,13,14,15,16,17,18,19,20,21,22}

Until now, only V or Cr doped (Bi,Sb)$_2$Te$_3$ systems have been experimentally reported to show QAH effect at an extremely low temperature ($<$100 mK), and the quantum anomalous Hall conductivity entirely vanishes at 2 K. \cite{23,24,25} In those doped topological insulators (TIs), magnetic disorder is found to be important to affect the temperature in detecting QAH effect. To acquire a perfect quantization of QAH effect in magnetically doped TIs system, a tremendously low temperature is hence needed to restrain magnetic disorders in experiments. \cite{26} Hence, the intrinsic magnetic topological materials are desired.

Experimentally, the honeycomb-Kagome (HK) structure of Y$_2$O$_3$ monolayer has synthesized on graphene.\cite{27} All the atoms are in the plane, which corresponds to high-symmetric $\emph{P6/mmm}$ structure. It indicates that the 2D HK structures could be prepared in the experiment. Wang $\emph{et al}$. firstly predict the HK structure of V$_2$O$_3$ monolayer has the ground state of FM state. More interestingly, the QAH effect is found to be realized in this system.\cite{28} Recently, Zhang $\emph{et al}$. predict the QAH effect can also be realized in 5d oxide Nb$_2$O$_3$ with the HK structure.\cite{29} These experimental and theoretical work indicate that the HK structure has a potential on the realization of the QAH effect.

In this paper, we have implemented systematic magnetic and electronic calculations on the basis of the density functional theory for 2D Mn$_2$O$_3$ monolayer. First of all, we find that the ground state is not in fact the high-symmetric $\emph{P6/mmm}$ structure, (the high-symmetric structure proposed in the previous studies \cite{27,28,29}), but a lower-symmetric phase with oxygen atoms shifting along the z axis. In addition, we also find that all Mn$_2$O$_3$ structure are ferromagnetic and Dirac half-metals (DHMs) with combination of a single-spin massless Dirac fermions and a half-metallic band structure. Finally, we find that the biaxial strain can drive the structure and topological phase transitions. These findings make Mn$_2$O$_3$ promising building blocks for future applications in electronics and spintronics.

\section{STRUCTURES AND COMPUTATIONAL METHODS}

To investigate the electronic and magnetic structures, we implemented the Vienna $Ab$ $initio$ Simulation Package (VASP) \cite{30,31} for the first-principles calculations based on density functional theory (DFT). The electron exchange-correlation functional was described by the generalized gradient approximation of the Perdew-Burke-Ernzerhof functional. \cite{32} The plane-wave basis set with a kinetic energy cutoff of 500 eV was employed. Here, $12\times 12\times 1$ ($12\times 8\times 1$) and $24\times 24\times 1$ ($24\times 16\times 1$) $\Gamma$-centered $k$ meshes are adopted for the structural optimization and the self-consistent calculations of the $\emph{P6/mmm}$ ($\emph{Cmm2}$, $\emph{C222}$) structure, respectively. To avoid unnecessary interactions between the monolayer, the vacuum layer was set to 20 \AA. The total energy convergence criterion was set to be 10$^{-6}$ eV. To confirm the structural stability, the phonon spectra were calculated using a finite displacement approach as implemented in the PHONOPY code, in which a $4\times 4\times 1$ supercell were used. \cite{33} What's more, to describe the strongly correlated 3d electrons of the Mn atom, the GGA + U method was used.\cite{34} The onsite Coulomb repulsion U was varied between 0 and 5 eV for Mn. An effective tight-binding Hamiltonian constructed from the maximally localized Wannier functions (MLWFs) was employed to explored the edge states. \cite{35,36,37} Therefore, the edge states were calculated in a half-infinite boundary condition using the iterative Green's function method by the package WANNIERTOOLS. \cite{37,38}

\section{RESULTS AND DISCUSSION }
\subsection{Search for the stable structure}
Figure 1(a) shows the structure with $\emph{P6/mmm}$ symmetry. The HK structure is that two Mn atoms form honeycomb lattice and three O atoms form kagome lattice. Evidently, 2D Mn$_2$O$_3$ with such a structure has the same $\emph{D$_{6h}$}$ point group symmetry as graphene. It is a planar single-layer sheet, and its optimized lattice parameter is a = b = 6.19 \AA.

In order to investigate dynamical stability of the 2D Mn$_2$O$_3$, we perform phonon spectrum calculations, as illustrated in Figure 2(a). It is well-known that the imaginary frequencies indicate the dynamical instability. The phonon spectrum of 2D Mn$_2$O$_3$ in $\emph{P6/mmm}$ phase indeed shows structural instabilities. These instabilities are indicated by the imaginary frequency modes. Note that $\Gamma$ point are doubly degenerate and M point are singly degenerate, and all modes with out-plane vibrational character. Particularly, we show the $\Gamma$ and M high-symmetry points eigenvectors in Figure 2(b). Thus, we search for the structural ground state by comparing the calculated total energies of structures with different oxygen atoms shift along z axis. $\emph{P6/mmm}$ space group have 24 symmetry operations, which consist of one identity operation ($\emph{E}$), one center of inversion operation ($\emph{I}$), one 180$^\circ$ rotation operation ($\emph{C$_2$}$), two 120$^\circ$ rotation operations ($\emph{C$_3$}$), two 60$^\circ$ rotation operations ($\emph{C$_6$}$), two 120$^\circ$ rotation times reflection operations ($\emph{S$_3$}$), two 60$^\circ$ rotation times reflection operations ($\emph{S$_6$}$), six rotation times inversion operations ($\emph{C$^\prime$}$ or $\emph{C$^{\prime\prime}$}$ ) and seven plane of reflection operations ($\sigma$). While oxygen atoms are shifted along z axis, the possible existed two subgroup structures are $\emph{Cmm2}$ and $\emph{C222}$, as shown in Figure 2(c).

\begin{figure}[htb]
\begin{center}
\includegraphics[angle=0,width=1.00\linewidth]{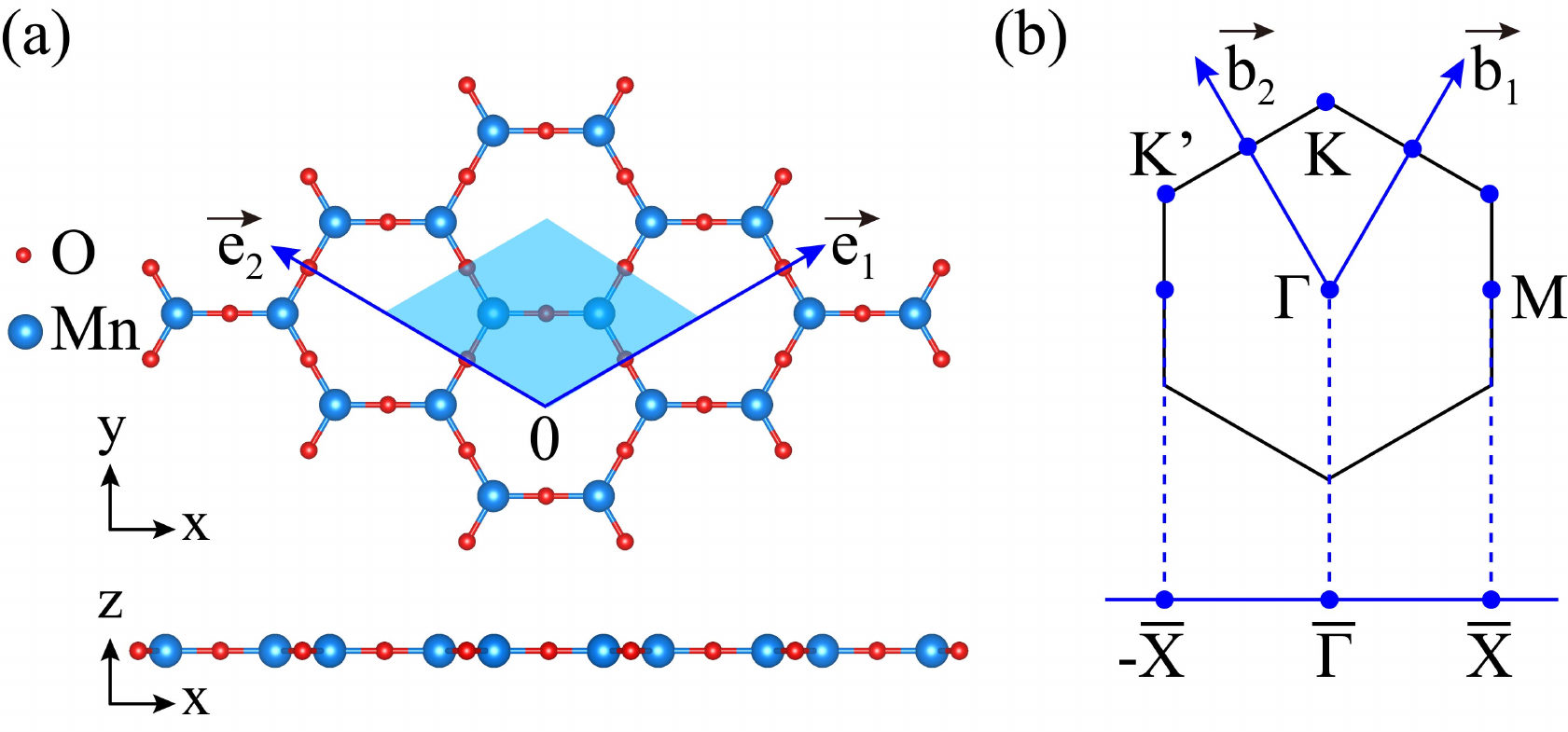}
\caption{(a) Top and side views of the $\emph{P6/mmm}$ structure of Mn$_2$O$_3$ monolayer with lattice vectors $\vec{e}_1$ and $\vec{e}_2$, the unit cell is indicated by the blue shading. The Mn and O atoms are depicted by the light blue and red balls, respectively. (b) The first Brillouin zone of the Mn$_2$O$_3$ with the reciprocal lattice vectors $\vec{b}_1$ and $\vec{b}_2$ and its projections onto the one-dimensional Brillouin zone.}
\end{center}
\end{figure}

\begin{figure}[htb]
\begin{center}
\includegraphics[angle=0,width=0.80\linewidth]{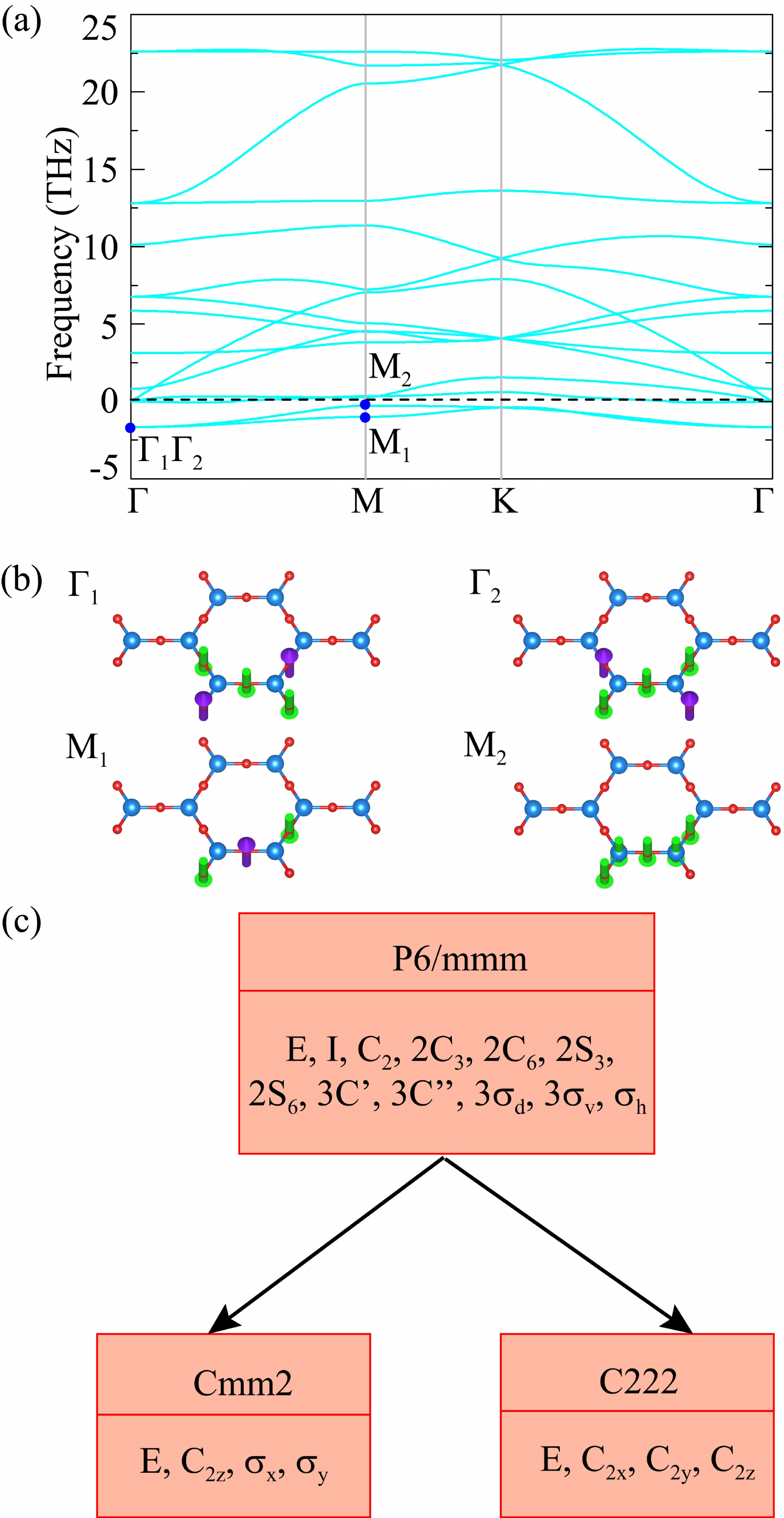}
\caption{(a) Phonon dispersion curves of the $\emph{P6/mmm}$ structure of Mn$_2$O$_3$ monolayer. (b) The imaginary wave numbers of the low-lying modes at Brillouin zone boundary $\Gamma$ and M points eigenvectors are shown in the lower panel. (c) A schematic diagram, indicating the possible 2 subgroup structures. Arrows indicate that a structure with unstable higher-symmetry to a stable lower-symmetry, which are the oxygen atoms along the z axis shift. }
\end{center}
\end{figure}

\begin{figure}[htb]
\begin{center}
\includegraphics[angle=0,width=1.00\linewidth]{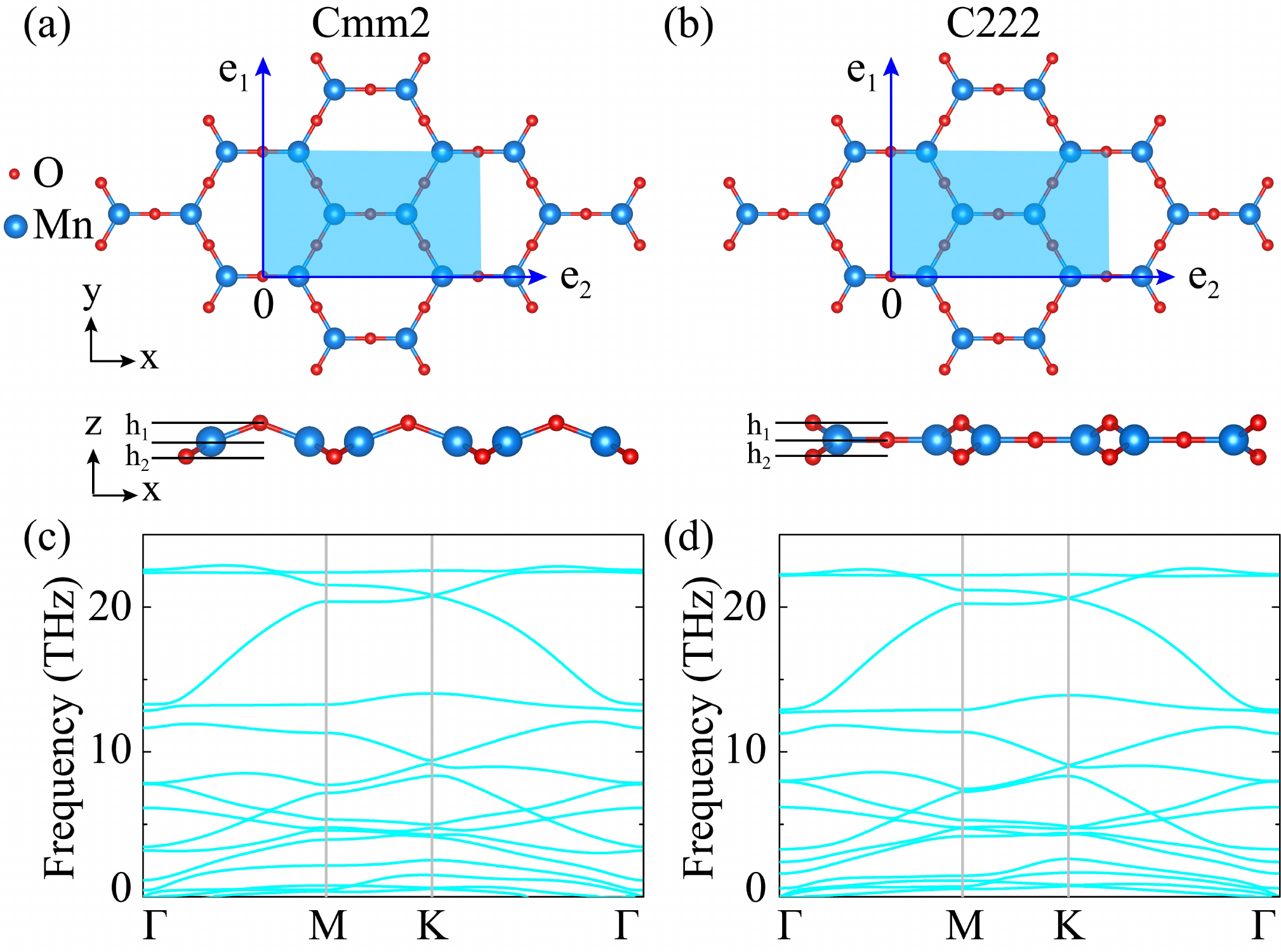}
\caption{(a, b) Top and side views of the $\emph{Cmm2}$ and $\emph{C222}$ structure of Mn$_2$O$_3$ monolayer with lattice vectors $\vec{e}_1$ and $\vec{e}_2$, the unit cell is indicated by the blue shading. The Mn and O atoms are depicted by the light blue and red balls, respectively. (c, d) Phonon dispersion curves of the $\emph{Cmm2}$ and $\emph{C222}$ structure of Mn$_2$O$_3$ monolayer, respectively.}
\end{center}
\end{figure}

The crystal structures of $\emph{Cmm2}$ and $\emph{C222}$ phase are shown in Figure 3(a) and Figure 3(b). The unit cell is made up of four Mn atoms and six oxygen atoms. Moreover, $\emph{Cmm2}$ and $\emph{C222}$ structure consist of free flat atomic layers. $\emph{Cmm2}$ structure only has four Mn atoms in its mid-plane, while the middle layer of $\emph{C222}$ structure has four Mn atoms and two oxygen atoms, and it is sandwiched by two oxygen atomic planes. h$_1$ (h$_2$) is the space between the top (bottom) layer of O and Mn atom layer, with values 0.26 $\rm \AA$ and 0.14 $\rm \AA$ for $\emph{Cmm2}$ phases and 0.23 $\rm \AA$ and 0.23 $\rm \AA$ for $\emph{C222}$ phase, respectively. We calculate the total energies of these two structures by fully relaxing the ionic positions. We find that the energy of $\emph{Cmm2}$ and $\emph{C222}$ structure is 4.65 meV and 4.54 meV per unit cell lower than that of $\emph{P6/mmm}$ structure, respectively. The energy differences between the $\emph{Cmm2}$ and $\emph{C222}$ phase are very small, under the range of numerical errors. As show in Figure 3(c), the phonon spectrum of $\emph{Cmm2}$ phase only has quite small imaginary frequencies in the vicinity of the $\Gamma$ point, which may be related to the limited supercell size. \cite{39} As shown in Figure 3(d), $\emph{C222}$ structure has no imaginary frequencies, showing the dynamically stability.

\subsection{The magnetic ground state}
In order to explore the magnetic ground state for three Mn$_2$O$_3$ phase, we will consider the contribution of three major energy: (1) magnetic energy interactions scale, (2) on-site Coulomb energy, (3) spin-orbit coupling energy. For magnetic energy, we consider all possible magnetic configurations in $2\times 2\times 1$ supercell (see Figure 4), namely, the FM, antiferromagnetic-Neel (AFM-N), antiferromagnetic-stripy (AFM-ST), antiferromagnetic-zigzag (AFM-ZZ), antiferromagnetic-cluster-I (AFM-C-I), antiferromagnetic-cluster-II (AFM-C-II), and antiferromagnetic-cluster-III (AFM-C-III). On the other hand, 3d Mn ions have a strong Coulomb correlation. We also calculate the electronic structure of Mn$_2$O$_3$ with different Hubbard U values. Finally, the spin-orbit coupling is considered. We calculate the total energy with the magnetization direction in or out-of the plane.

Figure 5(a) summarizes our calculated results. With nonzero Hubbard U values, the ground state of $\emph{P6/mmm}$ phase becomes FM$^z$ state, in which all the Mn magnetic moments align along the z axis (out of the plane). (all data, see Figure S1) Only when the Hubbard U disappears, the ground state changes to be antiferromagnetic with Mn magnetic moments in the plane (AFM-ST$^x$ configuration). Actually, the previous studies have shown that the on-site Coulomb repulsion U is 3$\sim$4 eV for Mn. \cite{40,41,42} Thus, combining $\emph{P6/mmm}$ phase and previous investigation, we studied the ground state of $\emph{Cmm2}$ and $\emph{C222}$ structure with U = 3 and 4 eV (all data, see Figure S2 and Figure S3), as illustrated in Figure 5(b, c). For all magnetic configurations, the total energies of FM$^z$ state are obviously smallest than other configurations. Hence, the ground state are FM$^z$ state for $\emph{Cmm2}$ and $\emph{C222}$ structure. In the following, we chose U= 4 eV to investigate band structure and topological properties.

\begin{figure}[htb]
\begin{center}
\includegraphics[angle=0,width=1.00\linewidth]{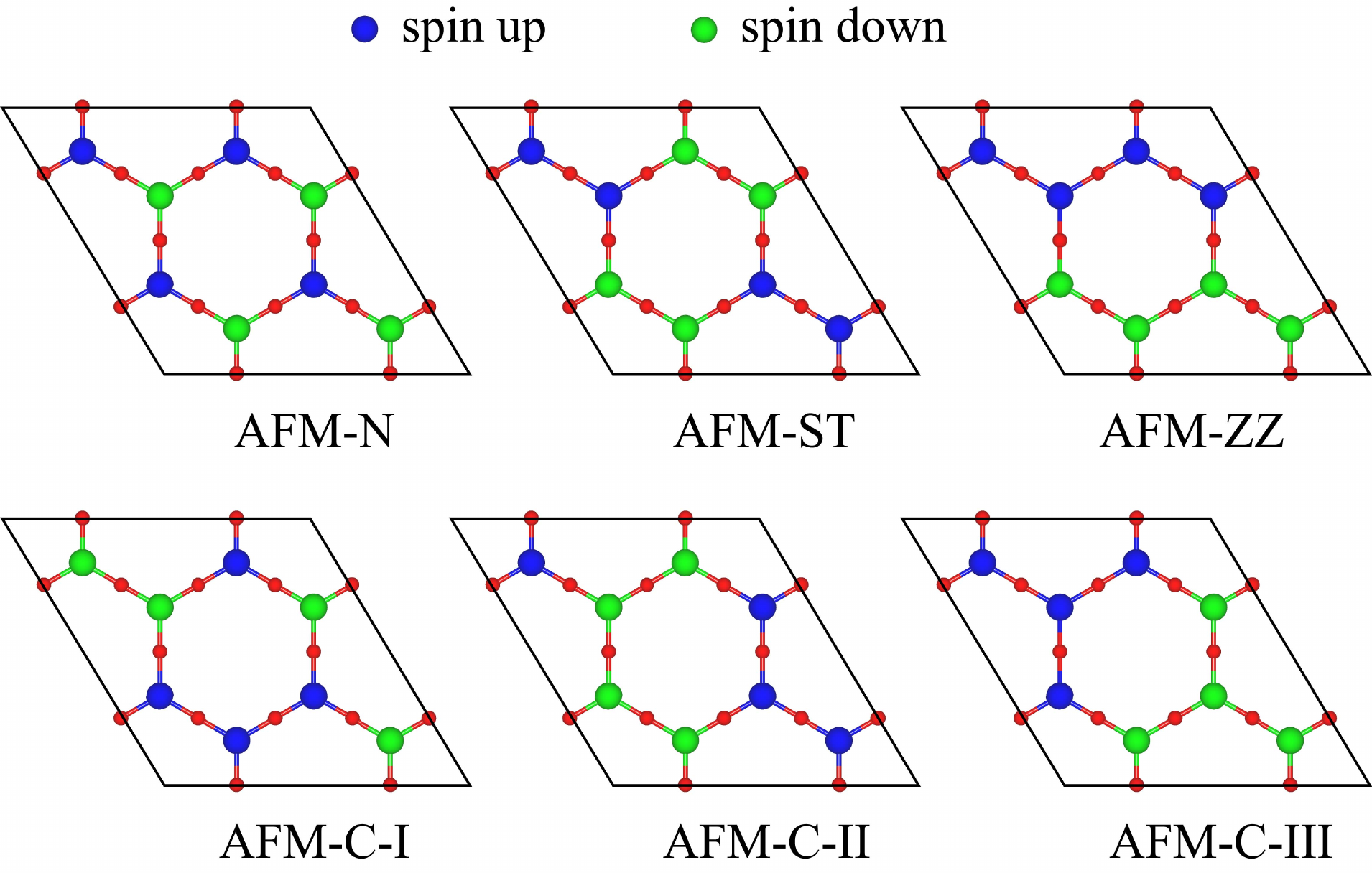}
\caption{Top view of various AFM spin configurations: AFM-Neel (AFM-N), AFM-stripy (AFM-ST), AFM-zigzag (AFM-ZZ), AFM-cluster-I (AFM-C-I), AFM-cluster-II (AFM-C-II), and AFM-cluster-III (AFM-C-III), where the blue (green) circles denote the up (down) spins.}
\end{center}
\end{figure}

Moreover, the spin polarized FM ground state with a magnetic moment of 4$\mu_B$ per Mn atom is found for Mn$_2$O$_3$. It corresponds to the d$^{4\uparrow}$ spin configuration of Mn$^{3+}$. The Monte Carlo (MC) method is used to calculate Curie temperature T$_c$. The MC simulations are implemented on a 80$\times$80 supercell which is adopted to reduce translational constraint, using 1$\times$10$^7$ loops for each temperature. T$_c$ value is obtained to be about 728 K, 606 K and 586 K for $\emph{P6/mmm}$, $\emph{Cmm2}$ and $\emph{C222}$, respectively.

\begin{figure}[htb]
\begin{center}
\includegraphics[angle=0,width=0.80\linewidth]{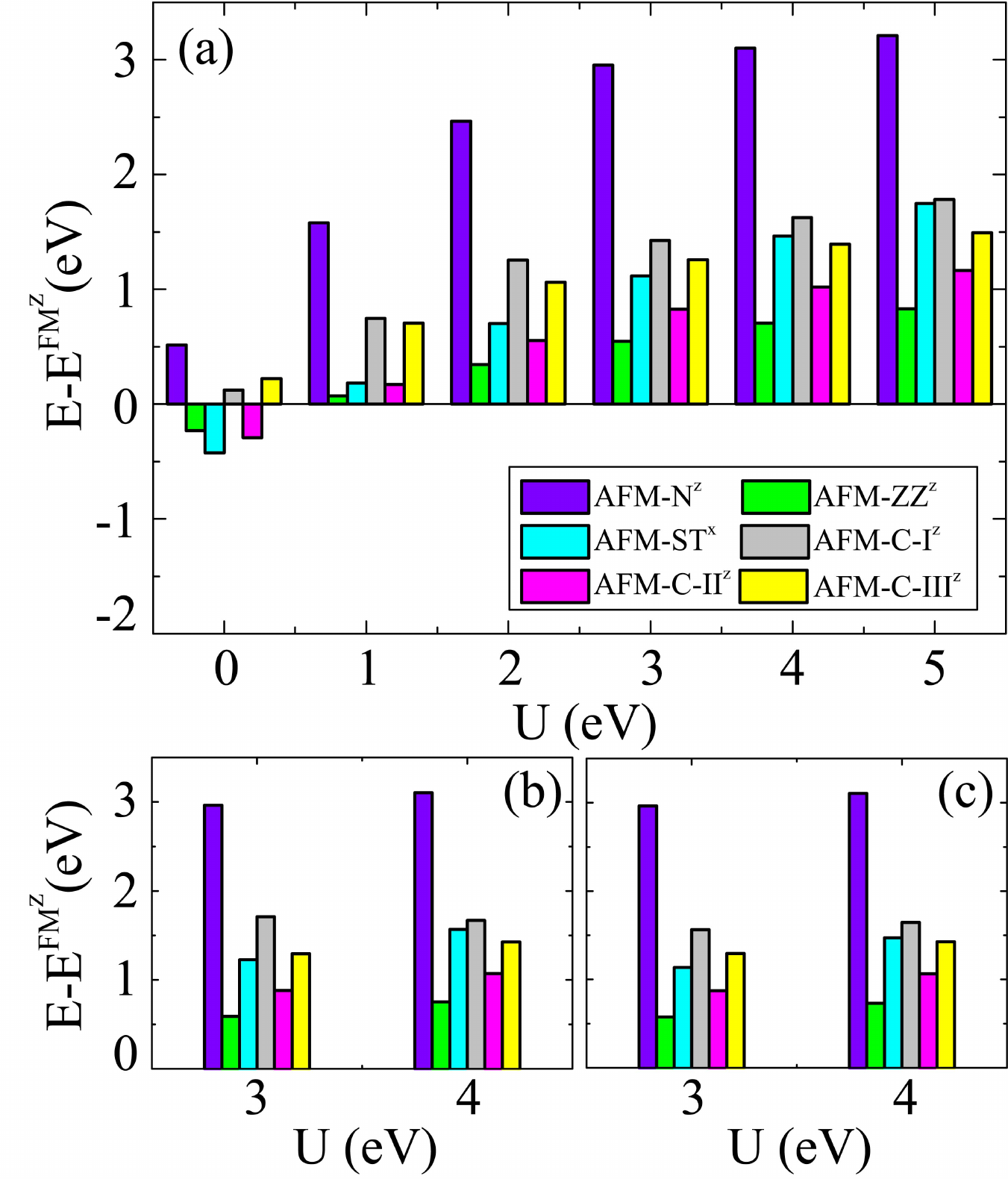}
\caption{The total energy E, relative to E of FM$^z$ state, for several magnetic configurations of Mn atoms in Figure 4 as a function of Hubbard U. (a) $\emph{P6/mmm}$ structure, (b) $\emph{Cmm2}$ structure and (c) $\emph{C222}$ structure. }
\end{center}
\end{figure}

\subsection{Electronic band structure}
In order to further investigate the nature of FM ordering in Mn$_2$O$_3$, we calculate the spin polarized band structures for $\emph{P6/mmm}$, $\emph{Cmm2}$ and $\emph{C222}$ phase, as shown in Figure 6 (a, d, g). We find that all structure exhibit two intriguing properties: (1) The spin down channel shows a large insulating gap with values 4.02 eV, 4.01 eV and 4.01 eV for the $\emph{P6/mmm}$, $\emph{Cmm2}$, and $\emph{C222}$ structure, respectively. (2) Four up-spin bands emerge near the Fermi level, two of which form Dirac cones at the high symmetry K point. In distorted $\emph{Cmm2}$ and $\emph{C222}$ structures, the Dirac cone in up-spin bands still exist but are inclined. It may be caused by the oxygen atoms shift along z axis. Therefore, the 2D Mn$_2$O$_3$ monolayer is an intrinsic DHM. To know the origin of the Dirac cone, we analyze the band contributions around the Fermi level, as shown in Figure S4. One can see that the Dirac states are mainly contributed by the d$_{xy}$ and d$_{x^2-y^2}$ orbitals of Mn atoms. These orbitals have stronger SOC strength than the $\emph{p}$ orbitals observed in the previous Dirac cones.\cite{2,3,21,43,44,45,46} The combination of massless Dirac fermion and 100$\%$ spin polarization render 2D Mn$_2$O$_3$ monolayer candidate for future applications in optoelectronics and spintronics.

\begin{figure}[htb]
\begin{center}
\includegraphics[angle=0,width=1.00\linewidth]{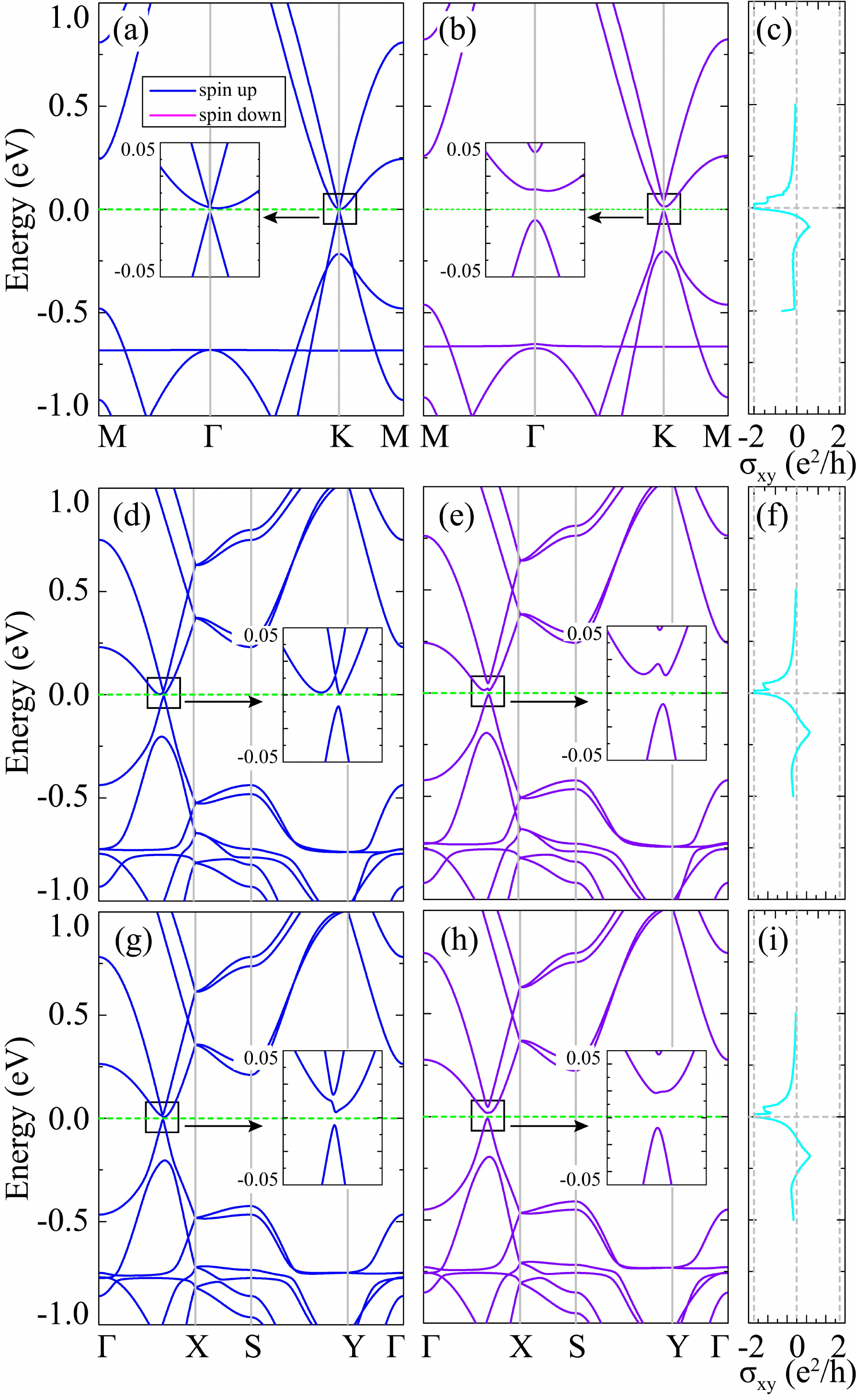}
\caption{Band structures without (a, d, g) and with (b, e, h) SOC as well as anomalous Hall conductivity ($\sigma_{xy}$) (c, f, i) of the $\emph{P6/mmm}$ structure (a, b, c), $\emph{Cmm2}$ structure (d, e, f) and $\emph{C222}$ structure (g, h, i). The blue and red curves correspond to the spin up and spin down energy bands, respectively.}
\end{center}
\end{figure}

Further, we study the influence of SOC on the band structures. Indeed, the Dirac band dispersion is destroyed and a band gap is opened when the SOC effect is considered. In Figure 6(b, e, h), we show the electronic structure with SOC, and band gaps of about 21.67 meV ($\emph{P6/mmm}$), 22.56 meV ($\emph{Cmm2}$) and 26.04 meV ($\emph{C222}$) are found.

\subsection{Quantum anomalous Hall effect}
To investigate the topological properties of Mn$_2$O$_3$, we calculate the anomalous Hall conductivity (AHC) by using the formula£º
\begin{equation}
\sigma_{xy} = C\frac{e^2}{h},
\end{equation}
\begin{equation}
C= \frac{1}{2\pi} \int_{BZ} d^2k ~\Omega(\textbf{k}),
\end{equation}
where C is Chern number. Nonzero C values indicate the quantized anomalous Hall conductance $\sigma_{xy}$. \cite{47} The calculated AHC of three different Mn$_2$O$_3$ structure are displayed in Figure 6. All Mn$_2$O$_3$ phases [Figure 6 (c, f, i)] show the quantized conductance, $\sigma_{xy}$ = -2e$^2$/h in the gap. It indicates that all the phase are QAH insulators with Chern number C = -2, showing the strong potential of Mn$_2$O$_3$ in the transport.

In addition,
\begin{equation}
\Omega(\textbf{k})=-\sum_{n}f_{n}\sum_{n\prime \neq n}\frac{2Im \left \langle \psi_{nk} \mid v_{x} \mid \psi_{n\prime k} \right \rangle \left \langle \psi_{n\prime k} \mid v_{y} \mid \psi_{nk} \right \rangle}{(E_{n\prime}-E_{n})^2},
\end{equation}
where $\Omega(\textbf{k})$ is the Berry curvature in reciprocal space, $v_{x}$ and $v_{y}$ are operator components along the x and y directions and $f_{n}=1$ for occupied bands. \cite{48} The Berry curvature distribution of the QAH state is displayed in Figure 7(b, d, f). One clearly sees that the main contribution of Berry curvature comes from the vicinity of the high symmetric K and K' points with the same sign for $\emph{P6/mmm}$ structure. The Berry curvature has the maximum values along the $\Gamma$-X direction for $\emph{Cmm2}$ and $\emph{C222}$ structure.

According to the  bulk edge correspondence, \cite{49} the absolute value of the non-zero Chern number is related to the number of nontrivial chiral edge states connecting the valence and the conduction bands. With an effective concept of principle layers, an iterative procedure to calculate Green's function for semi-infinite sheet of Mn$_2$O$_3$ is performed. \cite{50,51} The energy and momentum dependence of local density of states at the edge can be obtained from the imaginary part of the surface Green's function. The edge states are shown in Figure 7(a, c, e). It is clear that there are two gapless chiral edge states that emerge inside the bulk gap. The results are consistent with Chern number C = -2.

\begin{figure}[htb]
\begin{center}
\includegraphics[angle=0,width=1.00\linewidth]{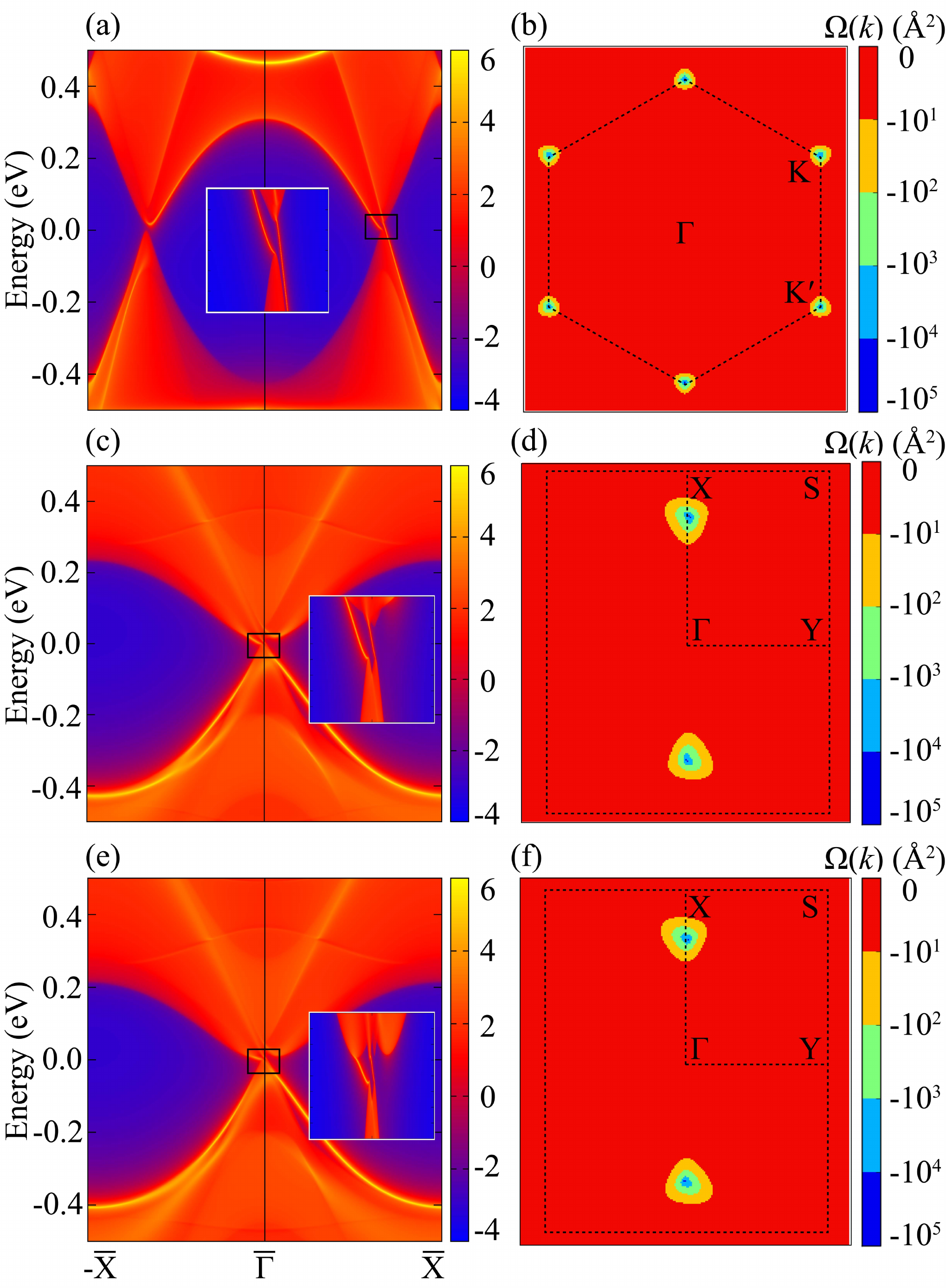}
\caption{(a, c, e) Calculated edge state of a semi-infinite sheet of Mn$_2$O$_3$, and (b, d, f) the Berry curvature with SOC in the momentum space of the $\emph{P6/mmm}$ structure (a, b), $\emph{Cmm2}$ structure (c, d) and $\emph{C222}$ structure (e, f).}
\end{center}
\end{figure}

\subsection{Strain driven structure and topological phase transitions}
Based on the calculations of the Chern number and edge states above, we have confirmed the nontrivial topology with sizable band gaps in three Mn$_2$O$_3$ structures. For the purpose of practical applications, it is of extraordinary interest to further explore the strain effect on the structure and topological properties. The applied strain can drive structural phase transitions in a series of materials, \cite{52,53,54} such as strain engineered T and H phase transitions in VS$_2$ monolayer. \cite{53} Besides this, the strain is expected to tune the band gap of our Mn$_2$O$_3$ system as well, and even change their band structure topology in a controllable way, which, if possible, will have potential use in electronics and spintronics. Therefore, we calculate electronic structures of three Mn$_2$O$_3$ phase under the biaxial strain to examine the tunability of a topological phase transition and the possible existence of a structures phase transition. The biaxial strain in our calculations is defined as $\varepsilon$ = (a-a$_0$)/a$_0$$\times$100$\%$, where a and a$_0$ represent the in-plane lattice constant after and before the strain is applied, respectively.

\begin{figure}[htb]
\begin{center}
\includegraphics[angle=0,width=0.70\linewidth]{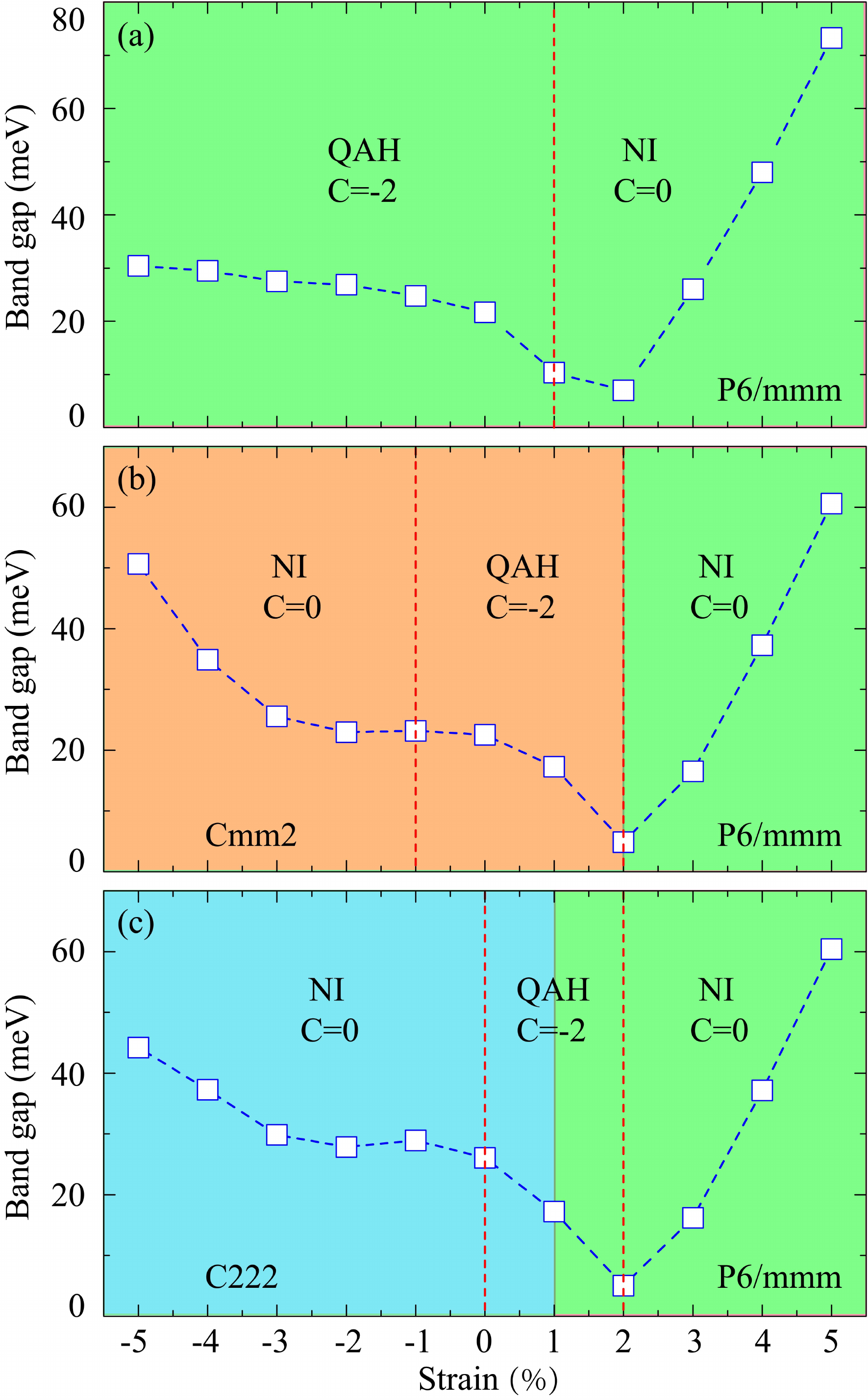}
\caption{Calculated global gap in three Mn$_2$O$_3$ monolayer structure as a function of the biaxial strain with U = 4 eV for Mn. Red vertical dashed lines denote the topological phase boundaries. Different background color denote different structure Mn$_2$O$_3$, the $\emph{P6/mmm}$, $\emph{Cmm2}$, and $\emph{C222}$ structure are depicted by the light green, orange, and light blue, respectively. (a) $\emph{P6/mmm}$ structure, (b) $\emph{Cmm2}$ structure and (c) $\emph{C222}$ structure.}
\end{center}
\end{figure}

As shown in Figure 8, we show the band gap, structural phase transitions and the topological phase diagram of the three Mn$_2$O$_3$ structure as a function of strain. The corresponding band structures of three Mn$_2$O$_3$ phase are illustrated in Figure S5, S6 and S7. It is seen that the variation of the band gap, structural phase and topological phase are all sensitive to the applied strain. We found that the nontrivial QAH insulating state is stable in a larger strain range until the strain reaches $\varepsilon$ = 1$\%$, where the $\emph{P6/mmm}$ structure becomes the normal insulator (NI). Moreover, Figure 8(a) shows that the global band gap can reach up to 30.4 meV in $\emph{P6/mmm}$ structure under a 5$\%$ in-plane extensive strain. Due to the $\emph{P6/mmm}$ phase is a planar structure, its structure is difficult to be changed under the applied strain. We also find that $\emph{Cmm2}$ and $\emph{C222}$ phase are located vastly close to the phase boundary between the QAH insulator phase and a normal insulator phase. Only within the a strain ranges from -1$\%$ to 2 $\%$ and 0$\%$ to 2$\%$ for $\emph{Cmm2}$ and $\emph{C222}$ structure, respectively, they remain in the QAH phase. The Dirac points are destroyed by tensile strain for $\emph{P6/mmm}$ phase and compressive strain for $\emph{Cmm2}$ and $\emph{C222}$ phase. Meanwhile, small strain will induce the structural transition from $\emph{Cmm2}$ to $\emph{P6/mmm}$ (2$\%$), and from $\emph{C222}$ to $\emph{P6/mmm}$ (1$\%$), respectively. Therefore, if SiC and h-BN are substrates experimentally, quantum devices in spintronics can be realized.

\section{CONCLUSION}
In summary, based on first principles calculations, we have systematically studied the structural stability, magnetic, and bands structures of the Mn$_2$O$_3$ monolayer. More interestingly, we find that the 2D Mn$_2$O$_3$ have three phases: $\emph{P6/mmm}$, $\emph{Cmm2}$ and $\emph{C222}$ phase, of which $\emph{Cmm2}$ and $\emph{C222}$ phase are more stable ones. All three phases are intrinsic Dirac half-metals which exhibits many fascinating properties, including massless Dirac fermions, 100$\%$ spin polarization, and large magnetic moments. Remarkably, a topological quantum phase transition between the QAH insulating phase and normal insulating phase can induced by strain. Meanwhile, the strain can also lead to the structural phase transition from $\emph{Cmm2}$ and $\emph{C222}$ phase to $\emph{P6/mmm}$ phase. The prediction of the QAH effect in the Mn$_2$O$_3$ monolayer provides a new platform with distorted structures for the exploration of QAH insulator in the transition metal oxide.

\section*{ACKNOWLEDGEMENTS}
This work was carried out at Lvliang Cloud Computing Center of China, and the calculations were performed on TianHe-2.

\end{document}